\documentstyle[12pt,moriond,psfig]{article}
\begin{document}
\heading{Inflow and outflows in NGC 2992/3}

\author{P.-A. Duc $^{1}$, E. Brinks $^{2}$, V. Springel $^{3}$, B. Pichardo $^{4}$} {$^{1}$ IoA, Cambridge, UK $^{2}$ Depto. de Astronom\'{\i}a, U. Guanajuato, Mexico} {$^{3}$ MPA, Garching, Germany $^{4}$ Instituto de Astronom\'{\i}a, UNAM, Mexico} 

\begin{moriondabstract}
We present observational evidences for inflow and outflows in the interacting system Arp 245 (NGC 2992/3).

\end{moriondabstract}

\section{Introduction}
 Whereas it is well known that gas may be driven into the core of merging disks, fueling
 a central AGN or a nuclear starburst, recent studies have shown that  a significant
fraction of the stellar/gaseous components is  expelled into the intergalactic
 medium along tidal tails (e.g., Hibbard \& van Gorkom, 1996; Duc et al, 1997). The tidal debris might 
be dispersed in the intergalactic/intracluster 
medium where it adds to the  diffuse background light such as that observed in the Coma cluster
(Gregg \& West, 1998), might fall back towards the merger, or regroup to form a new generation
 of galaxies,
the so-called  tidal dwarf galaxies (TDGs). We illustrate in this paper these various 
phenomena with a detailed multi-wavelength 
study of the interacting system NGC 2992/3 (Arp 245).

\section{Observations and simulations}

NGC 2992/3 (Arp 245) is a nearby  system composed of two interacting spiral galaxies: NGC 2992,
an edge-on Sa with a Seyfert 2 nucleus and NGC 2993 a face-on LINER.
Table~1 summarizes our observing campaign of Arp 245. An optical image shows  in
addition to well-defined stellar tidal tails which emanate from each disk resp. to the North and
 East, a more diffuse
30--kpc long bridge. At 21 cm, the neutral gaseous component shows a stunning morphology (see Fig.~1a).
 The HI VLA map 
exhibits a long gaseous tail escaping from NGC 2992 with a massive clump at its tip, a bridge between the
two galaxies with an orientation slightly different than in the optical, and a large 
 ring-like structure emanating from NGC 2993. The latter feature has a shape similar to the two other 
intergalactic HI rings known sofar: 
 the ring in Leo, in the M96 group (Schneider et al., 1989) and the ring near NGC 5291 
(Malphrus et al., 1997, Duc \& Mirabel, 1998). These two rings are however much larger.
Towards the nucleus of NGC 2992, HI is seen in absorption (Fig.~1b).
Our H$\alpha$ map (Fig.~2a) which traces the ionized gas shows  emission line regions in the disk of NGC 2993, 
in the inner regions of NGC 2992 as well as in the outer regions along filaments and 
 at the tip of the northern tail. 
 
We could reproduce the overall morphology of the system  using N-body + SPH
 numerical simulations of the collision. They indicate that the system is observed
 in the early stage of the interaction 
just  after the galaxies passed each other for the first time, about 100 Myrs after perigalacticon. 
The end-product of the collision will be a complete merger.
The time scales provided by the numerical simulations are useful to date the various phenomena 
observed in NGC 2992/3 and discussed herebelow.

\begin{table}[b]
\caption{Observation log}
\begin{tabular}{ll}
\hline
Optical broad-band imaging & ESO/NTT(EMMI)+ CFHT(MOS) \\
H$\alpha$ imaging & ESO/NTT(EMMI) \\
Optical long-slit spectroscopy & ESO/NTT(EMMI) \\
Near-infrared images & ESO/MPI 2.2m (IRAC 2) \\ 
HI mapping & VLA C-config \\
\hline
\end{tabular}
\end{table}

\begin{figure}[t]
\centerline{\psfig{file=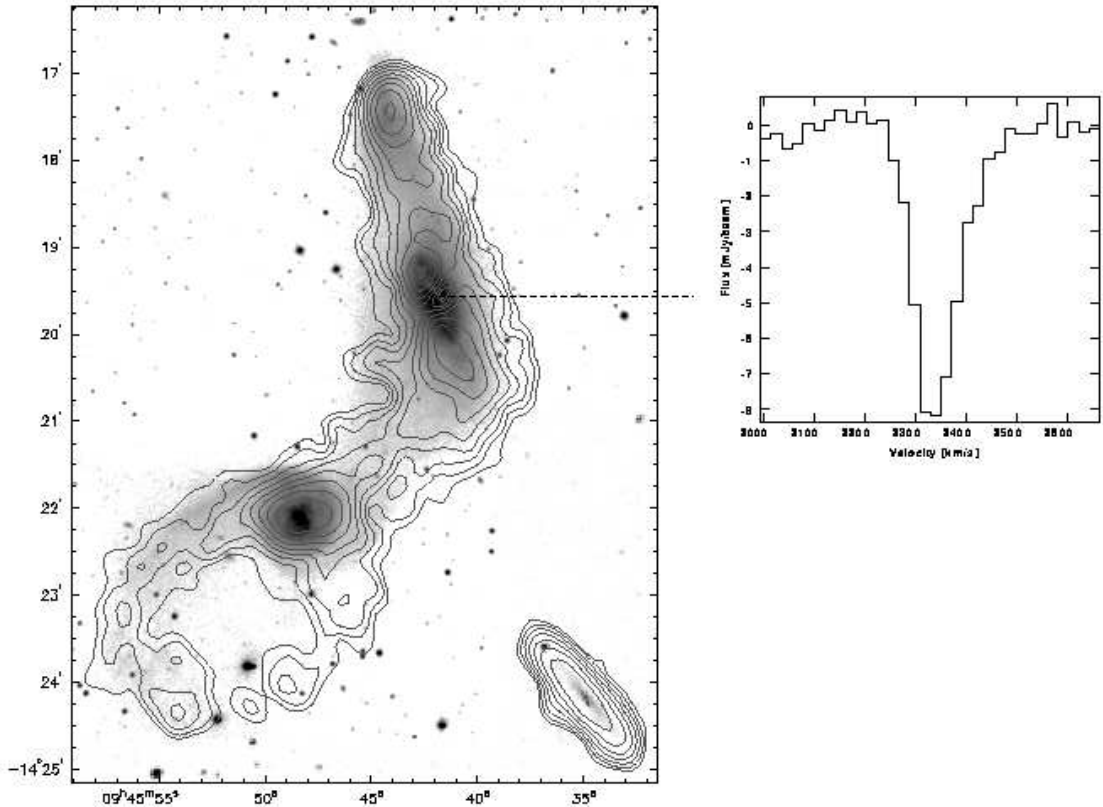,width=15cm,angle=-90}}
\caption{a) HI column density contours superimposed on an R-band image of the interacting system  
NGC 2992/3 (Arp 245). NGC 2992 is to the North, NGC 2993 to the South-East. The HI--rich object to the 
South-West is an edge-on disk at the same redshift.
 b) HI line spectrum towards the nucleus of NGC 2992.}
\end{figure}

\section{Inflow}
Galaxy collisions are efficient in driving gas towards the central regions. The mechanism 
involves the loss of angular momentum and gas transfer towards the central regions possibly via
a bar. This gas will then fuel a  starburst or an AGN. Suggestions of gas inflow
 in Arp 245 come from the analysis of the HI  spectrum towards the nucleus of NGC 2992.  
The line seen is absorption shows an asymmetry redwards of the systemic velocity which
is indicative of  foreground material falling towards the nucleus (see Fig.~1b). 
Whereas there is no evidence
for the presence of a nuclear starburst in NGC 2992,  strong Seyfert--2 type nuclear activity 
 is observed. Indeed, Seyfert--2 galaxies show a statistical excess of large companions with respect to
 nonactive disk galaxies (e.g: Dultzin-Hacyan et al., 1999).
One should note that the study of the inflow in NGC 2992 is largely 
hampered by the strong optical obscuration due to a prominent dust lane.

\section{Outflows}
Two kinds of outflows are observed in Arp 245 either in the form of small ionization filaments  
 or large tidal tails. These outflows have a
 different location and nature but have been directly or indirectly triggered by the
interaction.

\begin{figure}
\centerline{\psfig{file=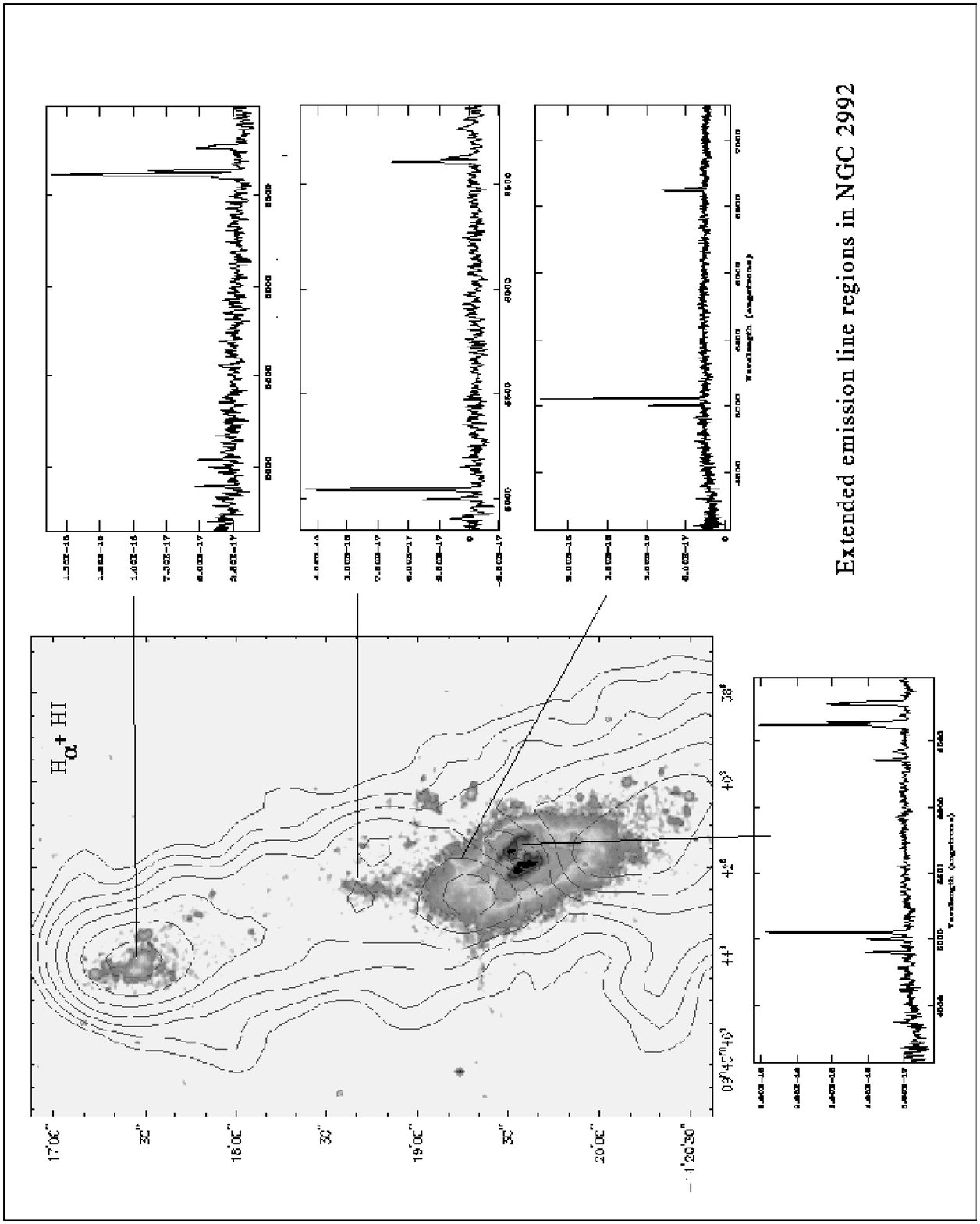,width=13cm,angle=-90}}
\caption{a) HI VLA contours superimposed on an H$\alpha$ image of NGC 2992. b) Optical
spectra of several emission line regions. The spectrum to the top is associated with a star--forming TDG
and the other ones with ionization filaments.}
\end{figure}

An ionization cone is found in the inner regions of NGC 2992 whereas at larger distances 
numerous extended emission line regions are observed escaping up to 10 kpc from the inclined disk of NGC 2992. 
They form narrow filaments that have relative radial velocities with respect to the surrounding HI gas 
 as high as 500 km/s (Marquez et al., 1998; Duc et al., in preparation). Their optical spectra 
 (see Fig.~2b) are consistent with a power-law type 
ionizing radiation source and a high ionization parameter reaching log(U)=-2 (see Fig.~2b). Outflowing
 gas could  be illuminated by a strong UV nuclear radiation. Other models invoke a strong local heating 
via shocks  (Allen et al.,1999) and gas dragged along by nuclear plasmoids jets.    

The material currently observed in the intergalactic medium, i.e along and at the tip of the bridge/tail/ring,
  was pulled out from the parent galaxies by tidal forces. 
The atomic hydrogen present in  these tidal features accounts for almost half of the total HI mass
 in the system (5.4 $10^{9}$~M$_\odot$). Such a distribution  is typical
of interacting systems that lose a significant fraction of their gas in the IGM. Part
of it might eventually fall back towards the parent galaxies or be tidally disrupted. 
 The gas clouds that are most likely to survive are those which are expelled to large distances
 and  are dense enough to become gravitationally bound, collapse and form stars.
An example of such an object, known as a ``Tidal Dwarf Galaxy'', is found at the tip of the
northern tidal tail which emanates from NGC 2992. At that location there is a $10^{9}$~M$\odot$ HI 
condensation associated with emission-line  regions (see Fig.~2a).
 Their optical spectra are typical  of HII regions ionized by
young massive stars (see Fig.~2b, up). The star-formation rate deduced from the H$\alpha$ luminosity is as high
 as 0.5~M$_\odot$~yr$^{-1}$. Nevertheless, broad band optical+near-infrared photometry of the tidal dwarf indicate 
 that its stellar population is still dominated at this stage of the collision by the old stars pulled out 
from the disk of the parent galaxy. 
 No star-forming regions appear to be associated with the HI ring south of NGC 2993.
One reason for that is that the HI there is not dense enough to have collapsed, contrary to the 
 tail of NGC 2992 or to the HI ring associated with
 NGC 5291 which hosts more than 20 TDGs (Duc \& Mirabel, 1998). A critical HI column density, 
 as high as $10^{21}$~cm$^{-2}$ in the case of Arp 245, is necessary for the onset of star
formation.\\

From spectra of the HII regions, we could estimate an 
 oxygen abundance of about solar for the TDG. This is higher than
the typical metallicity of TDGs (1/3 solar). TDGs are themselves more metal
rich than classical dwarf galaxies of the same luminosity (Duc \& Mirabel, 1998)
 due to the fact that they are recycled objects composed of already substantially enriched material.
In--situ enrichment in the TDG is unlikely given the constraint on the time scale  from the 
numerical simulations -- the interaction in Arp 245 is younger than 100 Myrs.
This implies that the material in the outer part parts of NGC 2992, which probably went into forming 
the TDG, was much more metal rich than gas in the outskirts of typical spirals where abundances of
 1/3 solar are usually measured.
A large scale enrichment  could have been induced by the nuclear outflows evidenced
 by the ionization filaments.
 Alternatively, the interaction has directly dragged material into the tidal tail which originally was 
close to the nucleus where solar metallicities are expected.\\

 
In any case, the inflow towards the nucleus and the outflows as exhibited by the
 ionization filaments and the tidal tails appear to be closely linked phenomena in 
Arp 245. They have either directly or  indirectly been triggered by the
collision.



%
\vfill
\end{document}